\documentclass[%
 reprint,
 superscriptaddress,
 amsmath,amssymb,
 prl
]{revtex4-2}

\usepackage{graphicx}
\usepackage{siunitx}
\usepackage{amsfonts}
\usepackage{mathtools}
\usepackage{gensymb}
\usepackage{bm}
\usepackage{amsmath}
\usepackage[colorlinks=true,bookmarks=false,citecolor=blue,linkcolor=blue,hyperfootnotes=true,urlcolor=blue]{hyperref}
\usepackage{nicefrac}

\DeclarePairedDelimiter\norm{\lVert}{\rVert}%
\begin{document} 

\title{Switchable X-ray Orbital Angular Momentum \\from an Artificial Spin Ice}

\author{Justin Woods}
\affiliation{Department of Physics and Astronomy, University of Kentucky, Lexington, KY 40506, USA}
\affiliation{Materials Science Division, Argonne National Laboratory, Lemont, IL 60439, USA}

\author{Xiaoqian Chen}
\affiliation{Advanced Light Source, Lawrence Berkeley National Laboratory, Berkeley, CA 94720, USA}
\affiliation{Department of Electrical and Computer Engineering, University of Kentucky, Lexington, KY 40506, USA}

\author{Rajesh V. Chopdekar}
\affiliation{Advanced Light Source, Lawrence Berkeley National Laboratory, Berkeley, CA 94720, USA}

\author{Barry Farmer}
\affiliation{Department of Physics and Astronomy, University of Kentucky, Lexington, KY 40506, USA}

\author{Claudio~Mazzoli}
\affiliation{National Synchrotron Light Source II, Brookhaven National Laboratory, Upton, New York 11973, USA}

\author{Roland Koch}
\affiliation{Advanced Light Source, Lawrence Berkeley National Laboratory, Berkeley, CA 94720, USA}

\author{Anton Tremsin}
\affiliation{Space Sciences Laboratory, University of California, Berkeley CA 94720, USA}

\author{Wen Hu}
\affiliation{National Synchrotron Light Source II, Brookhaven National Laboratory, Upton, New York 11973, USA}

\author{Andreas~Scholl}
\affiliation{Advanced Light Source, Lawrence Berkeley National Laboratory, Berkeley, CA 94720, USA}

\author{Steve Kevan}
\affiliation{Advanced Light Source, Lawrence Berkeley National Laboratory, Berkeley, CA 94720, USA}

\author{Stuart~Wilkins}
\affiliation{National Synchrotron Light Source II, Brookhaven National Laboratory, Upton, New York 11973, USA}

\author{Wai-Kwong Kwok}
\affiliation{Materials Science Division, Argonne National Laboratory, Lemont, IL 60439, USA}

\author{Lance E. De Long}
\affiliation{Department of Physics and Astronomy, University of Kentucky, Lexington, KY 40506, USA}

\author{Sujoy Roy}
\email{sroy@lbl.gov}
\affiliation{Advanced Light Source, Lawrence Berkeley National Laboratory, Berkeley, CA 94720, USA}

\author{J. Todd Hastings}
\email{todd.hastings@uky.edu}
\affiliation{Department of Electrical and Computer Engineering, University of Kentucky, Lexington, KY 40506, USA}

\begin{abstract}
Artificial spin ices (ASI) have been widely investigated as magnetic metamaterials with exotic properties governed by their geometries.  In parallel, interest in X-ray photon orbital angular momentum (OAM) has been rapidly growing.  Here we show that a square ASI with a programmed topological defect, a double edge dislocation, imparts OAM to scattered X-rays.  Unlike single dislocations, a double dislocation does not introduce magnetic frustration, and the ASI equilibrates to its antiferromagnetic (AF) ground state.  The topological charge of the defect differs with respect to the structural and magnetic order; thus, X-ray diffraction from the ASI produces photons with even and odd OAM quantum numbers at the structural and AF Bragg conditions, respectively. The magnetic transitions of the ASI allow the AF OAM beams to be switched on and off by modest variations of temperature and applied magnetic field. These results demonstrate ASIs can serve as metasurfaces for reconfigurable X-ray optics that could enable selective probes of electronic and magnetic properties.
\end{abstract}

\maketitle


\begin{figure}
    \includegraphics[width=0.95\linewidth]{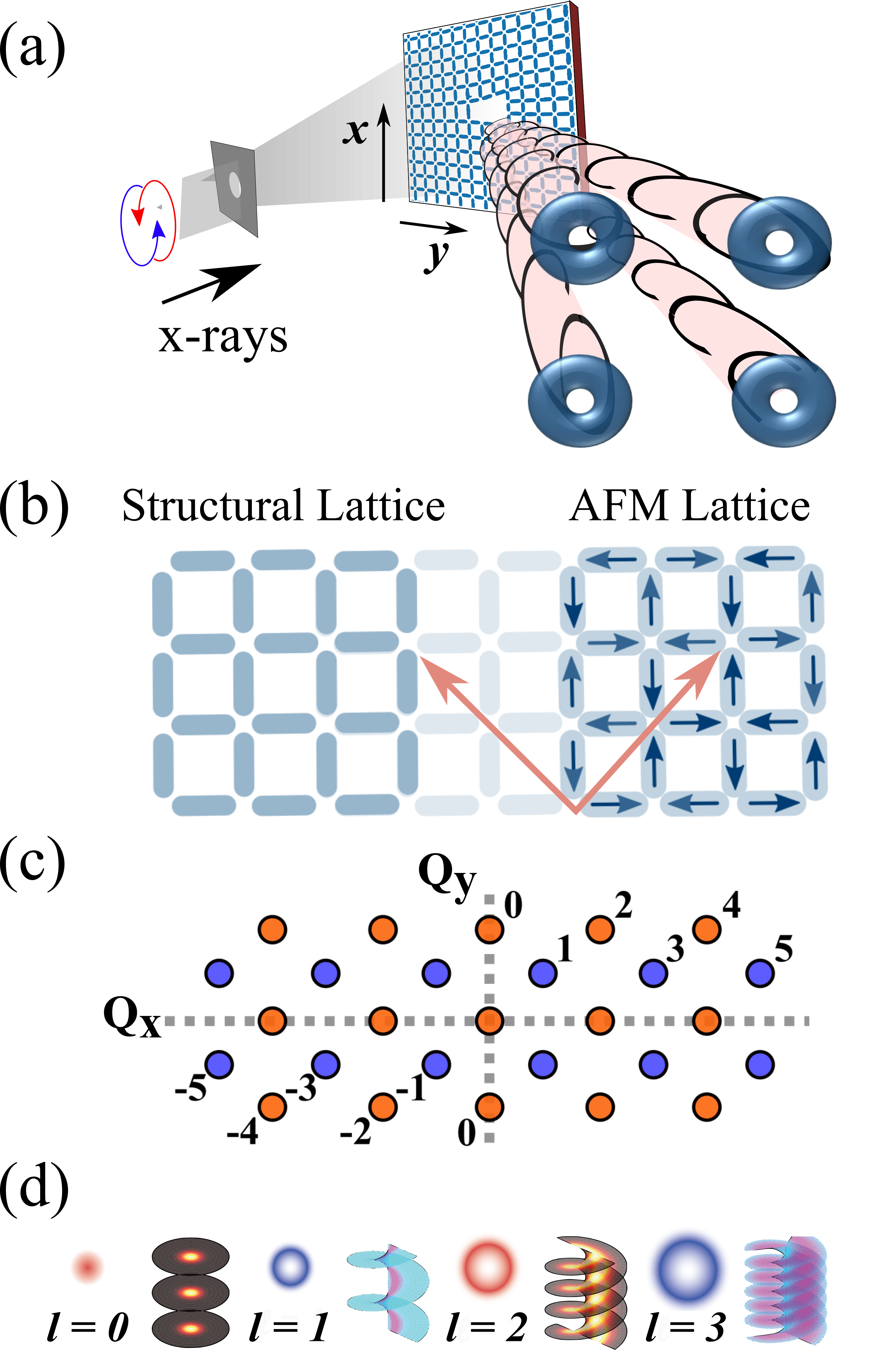}
    \caption{Schematic of X-ray OAM from a topological defect in an artificial spin ice.  (a) A square lattice of nanomagnets with a topological defect imparts orbital angular momentum to diffracted X-rays.  The structural-lattice defect has topological charge 2 and generates even-order OAM in the charge-scattered X-ray beams.  The antiferromagnetic ground state defect has topological charge 1 and produces odd-order OAM in the magnetically scattered X-ray beams (four lowest-order beams are shown). The magnetically scattered beams can be manipulated by varying  temperature or magnetic field. (b) Structural lattice and AF lattice. Red arrows show that the magnetic lattice is rotated $45\degree$ with respect to the structural lattice. (c) Positions of charge (red) and magnetic (blue) OAM peaks in reciprocal space labeled by OAM quantum number.  (d) Schematic of the intensity and phase fronts of beams with OAM quantum number $\ell= 0, 1, 2,$ and $3$. Red and blue represent charge and magnetic scattered beams carrying even- and odd-order OAM, respectively.}
    \label{fig:concept}
\end{figure}

Artificial spin ices (ASI) consist of patterned arrays of nanomagnets whose properties can be tuned based on geometry and competing interactions.  As a result, ASIs are often designed to realize systems not readily accessible in nature\cite{skjaervo_advances_2020,nisoli_colloquium:_2013} such as geometrically frustrated magnetic lattices\cite{perrin_extensive_2016,farhan_emergent_2019,gilbert_emergent_2016}.  ASIs have the advantage that they can be reconfigured through a variety of field\cite{farhan_direct_2013,sklenar_field-induced_2019}, temperature\cite{sendetskyi_continuous_2019,XiaoChenPRL}, and direct writing approaches\cite{wang_rewritable_2016, gartside_realization_2018}.  Applications of ASIs have recently begun to emerge with a primary focus on computing\cite{arava_computational_2018,arava_engineering_2019,iacocca_tailoring_2020,gliga_dynamics_2020,jensen_reservoir_2020}.  Here we consider a square ASI with a previously unstudied topological defect, a double edge dislocation, shown in Fig. \ref{fig:concept}(a).  Double edge dislocations are extremely rare defects in natural materials; however, they can be easily patterned into artificial systems.  We demonstrate that such a modified ASI can impart OAM to X-rays in a controlled manner.    

OAM is a topological property of light for which the photon phase has a helical structure around its propagation axis\cite{Allen,TerrizaReview,ShenReview}.  Interest in optical OAM is increasing in the X-ray regime where it could provide a selective probe of electronic and magnetic systems\cite{Loetgeringeaax8836,murnane,van_veenendaal_prediction_2007,van_veenendaal_interaction_2015,vila-comamala_characterization_2014,hemsing_coherent_2013,sakdinawat_soft-X-ray_2007,peele_observation_2002,cojoc_X-ray_2006}.  Recently, X-ray OAM was generated by gratings with structural topological defects\cite{lee_laguerregauss_2019}.  However, the phase properties of coherently scattered photons from a 2D-\textit{magnetic} lattice with a topological defect, as shown in Fig. \ref{fig:concept}(a), have not been studied in any detail.  In fact, there are only very limited reports of magneto-optic effects associated with OAM beams at any wavelength\cite{davis_intensity_1996,groshenko_optical_1998}. Here we employ tunable, coherent, X-ray sources to exploit the resonant enhancement of the spin-photon interaction cross section and observe soft X-ray OAM from ASIs.

To create an ASI with an edge dislocation one can displace the lattice points along one of the structural lattice vectors\cite{wilson_note_1950,willis_optical_1957}, taken as $\vec{\hat{x}}$ in Fig. \ref{fig:concept}(a), by a factor proportional to the azimuthal angle.\cite{S1}  The Burgers vector that describes the double dislocation is $\vec{t}=2a\vec{\hat{x}}$ where $a$ is the lattice constant.  Thus, two additional structural periods are acquired over the course of a Burgers circuit; consequently, the defect has structural topological charge $\mathbb{Z}_{s} = 2$, and we refer to this system as a ``Z2-ASI.'' Equivalently, the structural lattice acquires a geometric, or Pancharatnam–Berry, phase of $4\pi$ around a Burgers circuit.\cite{nye_dislocations_1974,cohen_geometric_2019} By choosing the proper lattice constant and material thickness, we hypothesized that the ASI would order into an AF ground state (See Fig. \ref{fig:concept}(b) and discussion below).  The AF lattice has twice the lattice constant of the structural lattice.  Thus, the magnetic topological charge of the defect would be $\mathbb{Z}_{m} = 1$, and the AF lattice would acquire a geometric phase of $2\pi$ around the Burgers circuit.  

We further predicted that X-ray diffraction from the Z2-ASI would yield odd and even OAM quantum numbers for magnetic- and charge-scattered beams at the AF and structural Bragg conditions, respectively as illustrated in Fig. \ref{fig:concept}(c).  Moreover, square ASIs exhibit antiferromagetic-to-paramagnetic (AF-to-PM) transitions with temperature \cite{kapaklis_melting_2012,XiaoChenPRL,sendetskyi_continuous_2019} and antiferromagnetic-to-ferromagnetic (AF-to-FM) transitions under applied fields\cite{farhan_direct_2013}.  The critical temperatures and reversal fields are governed by the size of and interactions among the nanomagnets along with the properties of the constituent material.  As a result, we predicted that the magnetically-scattered OAM beams should be sensitive to temperature and applied magnetic field.

\begin{figure*}
    \includegraphics[width=0.9\textwidth]{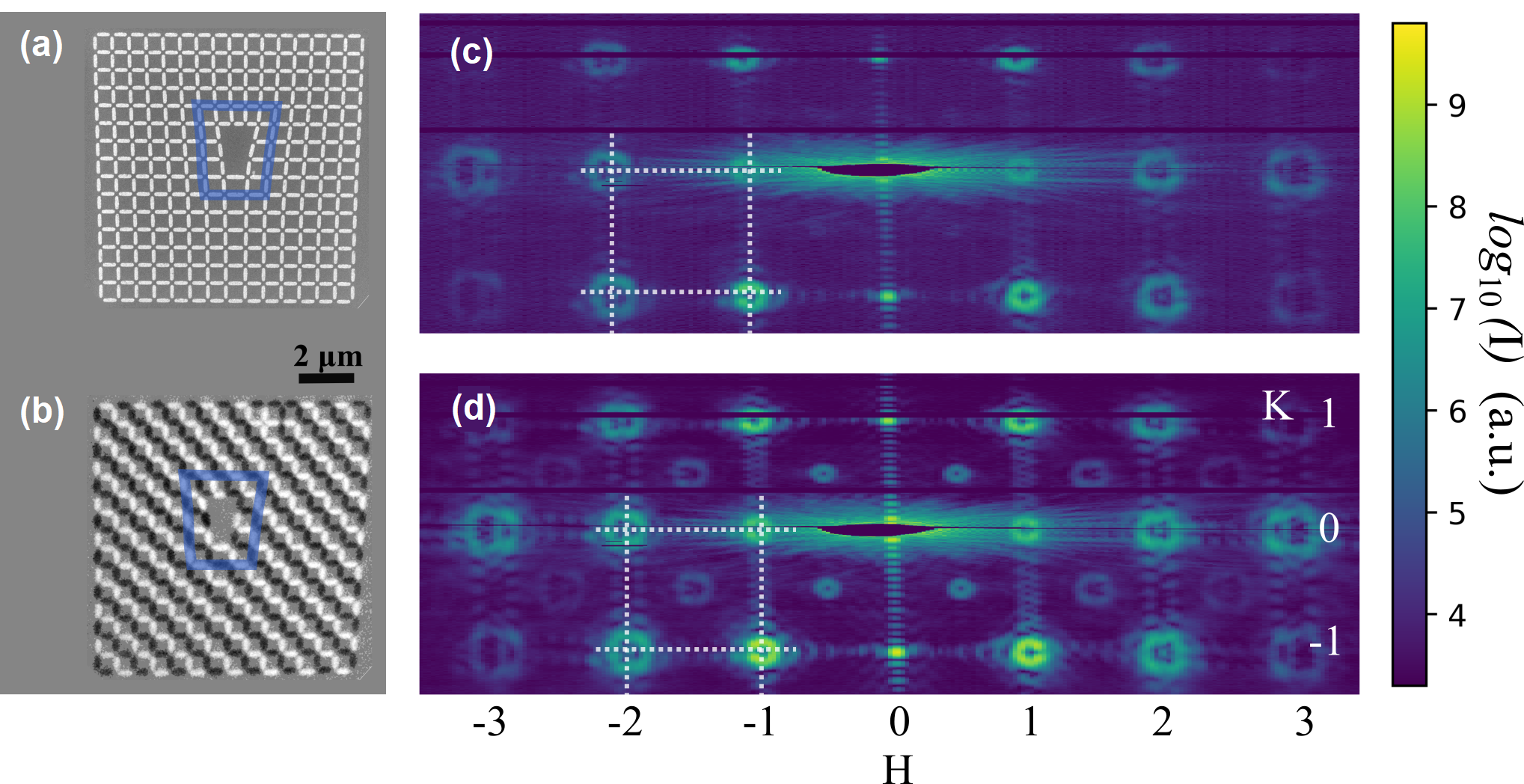}
    \caption{Experimental realization of a single-domain AF ground state in the Z2-ASI and resulting X-ray diffraction. (a) Scanning electron micrograph of a permalloy lattice of nanomagnets with a topological defect consisting of a double edge dislocation.  (b) X-ray magnetic circular dichroism PEEM micrograph revealing the AF ground state order. The blue boxes trace out a Burgers circuit.  (c) Off-resonance diffraction produces X-ray vortex beams carrying even order orbital angular momentum at the structural Bragg conditions. (d) Diffraction at the resonance condition dramatically enhances vortex beams at the AF Bragg conditions (half-integer values of $H$ and $K$) carrying odd-order OAM.}
    \label{fig:peem_and_OAMdiffraction}
\end{figure*}

To confirm these predictions, we fabricated the Z2-ASI from ferromagnetic permalloy ($\textrm{Ni}_{0.80}\textrm{Fe}_{0.20}$) using electron-beam lithography and lift-off.\cite{S1}  The resulting structure is shown in Fig. \ref{fig:peem_and_OAMdiffraction}(a).  The permalloy nanomagnets mimic Ising spins due to their large shape anisotropy. We chose the thickness of the permalloy to be between 2 and 3 nm so that the structure is thermally active and undergoes a AF-to-PM transition near room temperature ($T_N \approx$ 300 K). 

Similar structures without topological defects have been studied, and are known to order into an AF ground state.\cite{moller_artificial_2006,morgan_thermal_2011,zhang_crystallites_2013,farhan_direct_2013} The addition of a single dislocation introduces topological frustration and prevents the lattice from reaching a single-domain AF state\cite{drisko_topological_2017}.  However, for the double dislocation studied here, the magnetic lattice is not frustrated at the nearest-neighbor level.  Thus, thermal fluctuations near the AF-to-PM phase transition should enable the sample to attain a single-domain AF ground state when slowly cooled below $T_N$.\cite{XiaoChenPRL} However, it was not known \textit{a priori} whether the lattice distortion around the double edge dislocation would nucleate and/or pin superdomain walls or introduce longer nanomagnets that may not be thermally active\cite{S1}.  These effects could prevent long-range ground state ordering despite the absence of direct magnetic frustration.

To resolve these questions, we used X-ray magnetic circular dicrhroism photoemission electron microscopy (XMCD-PEEM) to experimentally image the ground state order as shown in Fig.~\ref{fig:peem_and_OAMdiffraction}(b).  XMCD-PEEM is a magnetic imaging technique that probes magnetization parallel or antiparallel to the incident X-ray beam. Magnetic images were recorded using the PEEM-3 microscope at Beamline 11.0.1.1 of the Advanced Light Source.  The incident X-rays were tuned to the Fe $L_3$ absorption edge ($\sim$708 eV), and the difference between the right- and left- circularly polarized X-ray images was taken to emphasize magnetic contrast over chemical or topographic contrast. 

The image in Fig.~\ref{fig:peem_and_OAMdiffraction}(b) was taken at $T = 110$ K, well below the AF-to-PM transition temperature.  The X-ray beam was oriented along the $[-1~1]$ direction of the structural lattice at 30$\degree$ grazing incidence.  This sample orientation gives sensitivity to in-plane magnetic moments for both $[1~0]$ and $[0~1]$ oriented islands in the lattice.  The white and black regions signify opposite magnetization directions of the nanomagnets, and confirm that the sample orders into an AF ground state. A Burgers circuit drawn around the defect, indicated in blue in Fig.~\ref{fig:peem_and_OAMdiffraction}(a) and (b), shows that the structural lattice has two extra periods ($\mathbb{Z}_{s} = 2$) while the magnetic lattice has one extra period ($\mathbb{Z}_{m} = 1$) as discussed above.  Thus, the  engineered defect provides the desired topological charge and does not prevent single-domain ground state ordering.

With the antiferromagnetic ground state established, we conducted coherent X-ray scattering experiments to study the OAM of the diffracted beams.  X-ray scattering was conducted at the COSMIC-Scattering beamline at the Advanced Light Source, Lawrence Berkeley National Laboratory, and at the Coherent Soft X-ray (CSX) beamline at the National Sychrotron Light Source II, Brookhaven National Laboratory. The diffracted photons were detected using a Timepix\cite{tremsin_overview_2020,tremsin_optimization_2018,fiorini_single-photon_2018} based soft X-ray detector at ALS and fast CCD detector at CSX. The diffraction experiment was performed in a reflection geometry as shown in Fig.~\ref{fig:concept}(a). A coherent beam of X-rays was incident on the sample at $\theta=9\degree$.

To obtain magnetic sensitivity, we employed resonant magnetic scattering by tuning the energy of the incident X-ray to Fe $L_3$ edge (708~eV).  Figures~\ref{fig:peem_and_OAMdiffraction} (c) and (d) show the diffraction patterns from the Z2-ASI with X-rays tuned away from (690 eV) and on (708 eV) the Fe $L_3$ absorption edge, respectively.  Vortex beams, bright rings with zero intensity at the center, appear at the structural Bragg condition, integral values of $(H,K)=\left(Q_x/(2\pi/a),Q_y/(2\pi/a)\right)$, for both off- and on-resonant X-ray illumination. On-resonance, additional vortex beams become visible due to magnetic scattering at the charge-forbidden (half-integer) AF Bragg condition (Fig.~\ref{fig:peem_and_OAMdiffraction}(d)).  Magnetic peaks have been previously observed from the AF ground state of square lattices without topological defects\cite{Perron_PRB_2013,XiaoChenPRL}, but here the topological defect yields vortex beams at the AF Bragg condition. 

The detailed characteristics of these diffracted X-ray beams are determined by the structure factor and form factor of the system.  However, the important information about the photon OAM can be found simply from the lattice sum, which, for a 2D square lattice with an edge dislocation, is given by
\begin{align}
    L&=\sum_m \exp{(i \vec{Q}\cdot \vec{R_m}+i \vec{Q}\cdot \vec{t} \: \frac{\psi_m}{2\pi})}
    \label{eq:latsum}
\end{align}
where $\vec{R_m}$ are the lattice vectors of the undistorted lattice, $\vec{t}$ is the Burgers vector for the structural lattice, and $\psi_m$ is the azimuthal angle for the $m^{th}$ lattice vector\cite{wilson_note_1950}.  Near a reciprocal lattice point, $\vec{Q_0}$, this sum can be approximated as\cite{S2}
\begin{align}
    L'\left(\rho', \phi'\right)=\displaystyle i^\ell \exp(i\ell\phi') U\left(\ell,\rho'\right) \label{eq:latticesum}.
\end{align}

The diffracted beam resulting from the lattice sum in Eq. \ref{eq:latticesum} is described by a phase factor, $\exp(i\ell\phi')$, and a purely real amplitude, $U(\ell,\rho')$.  The phase factor reveals that any beam with $\ell\neq 0$ carries an orbital angular momentum of $\hbar \ell$\cite{Allen} where the integer $\ell$ is given by
\begin{align}
    \ell=\frac{\vec{Q_0}\cdot\vec{t}}{2\pi}.
    \label{Eq:OAMQN}
\end{align}
As noted above, charge-scattered beams occur at integral valued reciprocal lattice vectors.  For the case of $\vec{t}=2a\vec{\hat{x}}$ considered here, Eq. \ref{Eq:OAMQN} yields $\ell = 2H$ where $H=Q_x/2\pi=1, 2, 3, ...$ and $\ell$ is always even.  Antiferromagnetically scattered beams occur at half-integer reciprocal lattice vectors where $H=Q_x/2\pi=1/2, 3/2, 5/2, ...$, and $\ell$ is odd.   Thus, the diffracted beams from the structural lattice carry even-order angular momentum (i.e. $0\hbar$, $2\hbar$, $4\hbar$,\dots) consistent with $\mathbb{Z}_{s}=\mathbb{N}=2$.  The magnetically diffracted beams carry odd-order angular momentum (i.e. $1\hbar$, $3\hbar$, $5\hbar$, \dots). This is consistent with $\mathbb{Z}_{m}=1$ because scattering with even-order angular momentum is forbidden by the symmetry of the AF structure factor.  The amplitude in Eq. \ref{eq:latticesum}, $U(\ell,\rho')$, reveals that the beams that carry OAM ($\ell \neq 0$) will exhibit a vortex structure.  The amplitude also indicates that the vortex radius will increase with $\ell$.\cite{S2} All of these results are consistent with the diffraction patterns shown in Fig.~\ref{fig:peem_and_OAMdiffraction} (c) and (d).
 
\begin{figure}
    \includegraphics[width=\linewidth]{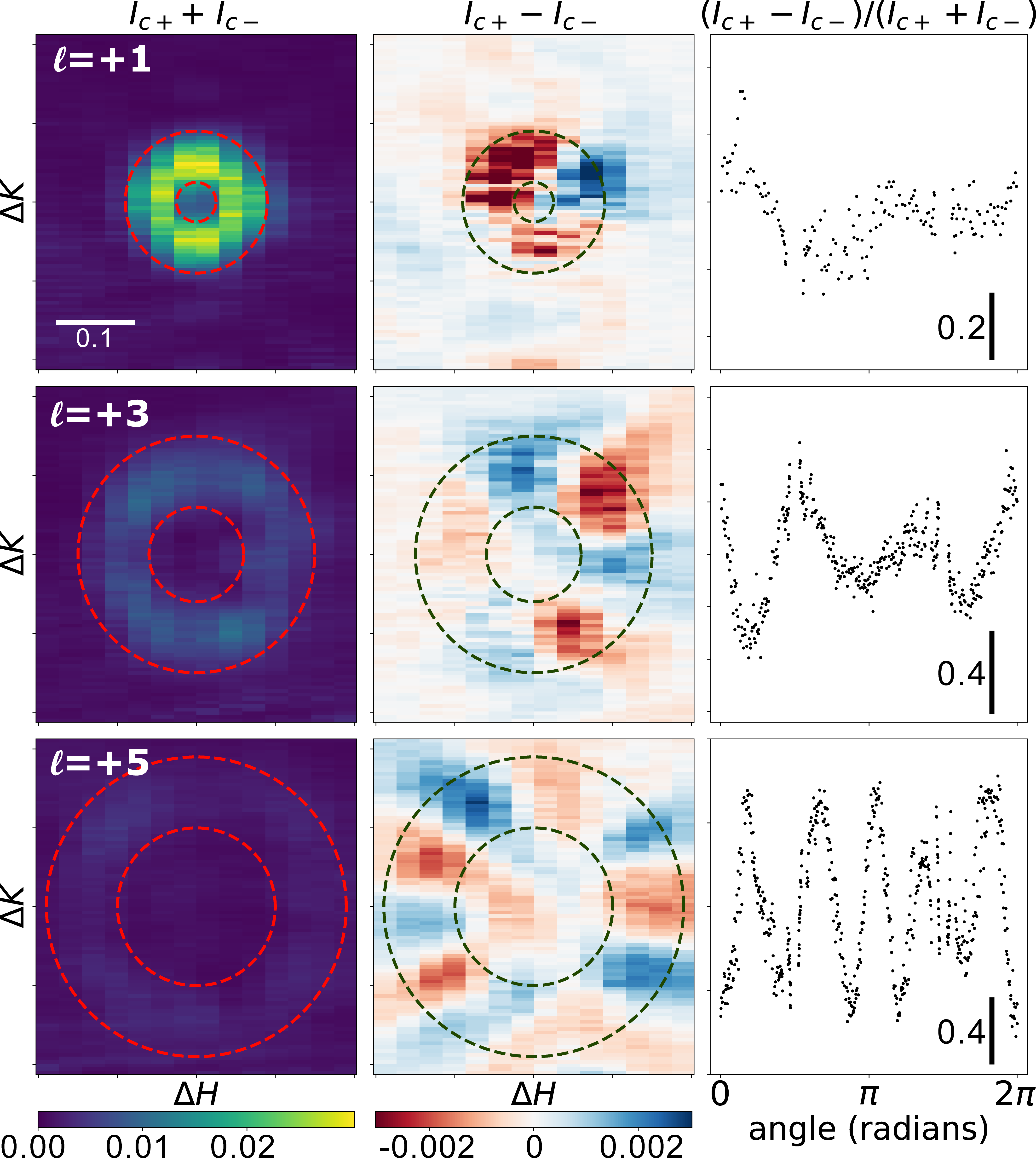}
    \caption{Determination of the AF OAM quantum number from interference between charge and magnetically scattered X-rays. The sum and difference of diffracted intensities (arb. units) from left- ($I_{c+}$) and right- ($I_{c-}$) circularly polarized illumination are plotted for magnetic beams with $\ell=+1$, $\ell=+3$, and $\ell=+5$ OAM quantum numbers.  The first column, $I_{c+}+I_{c-}$, shows the vortex structure of the beams.  The second column, $I_{c+}-I_{c-}$, reveals the interference pattern between charge and magnetically scattered X-rays.  Dashed lines serve as a guide and the scale bar applies to all panels.  Within the vortex ring, the number of interference fringes corresponds to the OAM quantum number as it would for interference with a plane wave.  The third column plots ${I_{c+}-I_{c-}}/{I_{c+}+I_{c-}}$ versus azimuthal angle which reveals an approximately sinusoidal oscillation consistent with the OAM phase progression.}
    \label{fig:interference}
\end{figure}

The polarization sensitivity of X-ray scattering\cite{hill_x-ray_1996} also enables direct determination of the phase progression of the OAM beams and thus the OAM quantum numbers.  Specifically, $\sigma$-polarized incident X-rays yield charge scattered beams carrying even OAM with $\sigma$-polarization and magnetically scattered beams carrying odd OAM with $\pi$-polarization.\cite{S3}  For circularly polarized incident light, there is interference between the charge and magnetically scattered components.  These interference effects are most easily observed in the difference in scattered intensity between left- and right- ($I_{c+}$ and $I_{c-}$) circularly polarized illumination, as shown in Fig.~\ref{fig:interference}.  The circular-polarization sensitivity arises from the phase difference in the $\pi$ component of left- and right- circularly polarized incident light.  The interference patterns in Fig.~\ref{fig:interference} exhibit modulation of the intensity in the azimuthal direction about the axis of propagation.  This is a result of the azimuthal phase progression of the primary beam with respect to the slow (or zero) phase progression in the tails of nearby charge-scattered beams.  The increasing number of fringes with diffracted order is consistent with increasing OAM and thus an increasing number of phase windings.\cite{white_interferometric_1991,pan_measuring_2018}  Plotting $I_{c+}-I_{c-}$ vs. azimuthal angle for the first three AF beams gives the number of fringes as $\ell = $1, 3, and 5 which is the same as the predicted OAM quantum number from Eq.~\ref{Eq:OAMQN}.

\begin{figure}
    \includegraphics[width=\linewidth]{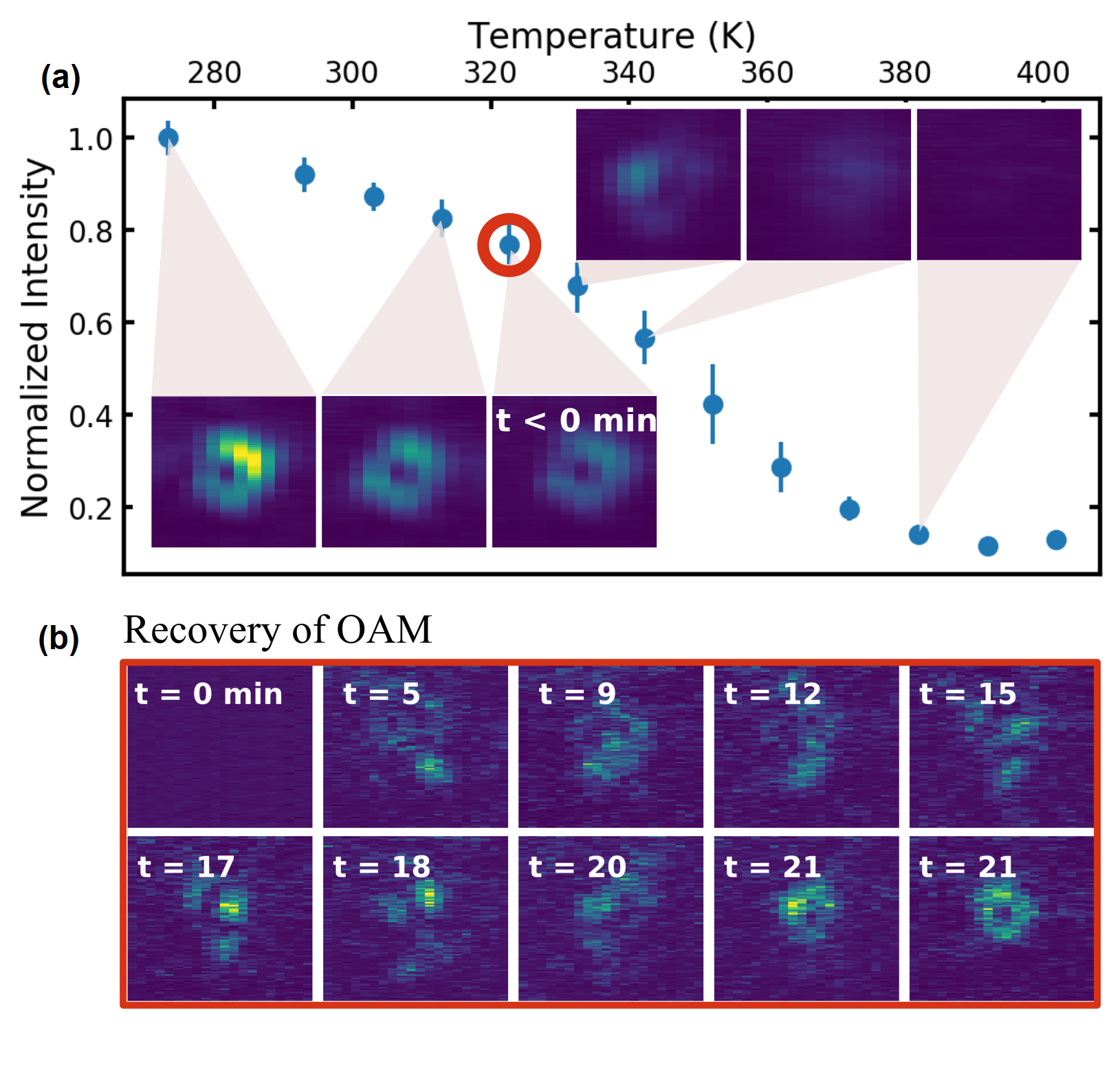}
    \caption{Response of the Z2-ASI to temperature and applied magnetic field.  (a) Temperature dependent X-ray intensity of the magnetically scattered OAM beam with $\ell = 1$. As the sample approaches the antiferromagnetic-to-paramagnetic transition temperature ($T_N \approx 380$ K) of the artificial lattice, the integrated intensity near the AF Bragg condition decreases.  (b) Time dependence of intensity after the beams are switched off using a magnetic field.  The field is removed at $t = 0$ and the temperature is held at 320 K.  The intensity for $t<0$ is labeled in (a).  The speckle pattern at early time points indicates that disordered AF domains form first.  Afterwards, the sample further relaxes to its AF ground state and the OAM vortex beam is restored. Indicated times are in minutes.}
    \label{fig:temp_field}
\end{figure}

Finally, the temperature and magnetic field dependence of the AF order provides control of the AF-scattered OAM beams. Similar permalloy square lattices without topological defects undergo AF-to-PM phase transitions near room temperature\cite{XiaoChenPRL} and AF-to-FM transitions under applied magnetic field\cite{farhan_direct_2013}.  As temperature increases toward the AF-to-PM transition of our Z2-ASI, the intensities of the OAM-beams at the AF Bragg condition weaken until they are extinguished above $T_N \approx$ 380 K as shown in Fig.~\ref{fig:temp_field}(a).  This result was replicated using a sample with lower $T_N$\cite{S4}.

If we apply an in-plane magnetic field to the sample, instituting an AF-to-FM transition, the AF OAM beams are also extinguished, as shown in the first frame of Fig.~\ref{fig:temp_field}(b). In this case, we applied the field along the $[1~0]$ direction which was orthogonal to the beam axis.  If we maintain the sample temperature close to $T_N$ and remove the magnetic field, the sample relaxes into disordered AF superdomains within seconds as evidenced by the speckle patterns at short time scales of Fig.~\ref{fig:temp_field}(b).  The sample relaxes to the AF ground state over several minutes, as shown in the last frame of Fig.~\ref{fig:temp_field}(b). A video of the fluctuation and relaxation processes is available online\cite{S5}.  Thus, the AF OAM beams can be switched on and off with small changes in temperature and applied field.  

We have shown that introducing a double edge dislocation in a square artificial spin ice does not prevent equilibration to a single-domain antiferromagnetic ground state.  Thus, the programmed topological defect exhibits a structural charge of two and a magnetic charge of one.  As a result, the ASI imparts even- and odd-order orbital angular momentum to charge- and magnetically-scattered X-ray photons respectively.  We determined the OAM quantum numbers of the magnetically scattered beams using a novel self-interference technique which exploits the polarization dependence of the resonant X-ray scattering.  Finally, the ASI's AF-to-FM and AF-to-PM transitions allowed the AF OAM beams to be switched on- and off- with modest changes in field and temperature respectively.  These findings represent a first step toward realizing reconfigurable optics for the generation and analysis of soft X-ray orbital angular momentum.  More broadly, these studies show that engineering defects in nanoscale magnetic lattices offers a powerful tool for designing X-ray metasurfaces.

The work is supported by the U.S. Department of Energy, Office of Science, Office of Basic Energy Sciences under Award Number DE-SC0016519.  The work used Timepix based soft X-ray detector, development of which is supported by DOE through award RoyTimepixDetector. XMC and SDK acknowledge partial support through DOE BES DE-AC02-05-CH11231 within the Nonequilibrium Magnetic Materials Program (MSMAG). This research used resources of the Advanced Light Source, which is a DOE Office of Science User Facility under contract no. DE-AC02-05CH11231.  This research used the 23-ID-1 (CSX) beamline of the National Synchrotron Light Source II, a U.S. Department of Energy (DOE) Office of Science User Facility operated for the DOE Office of Science by Brookhaven National Laboratory under Contract No. DE-SC0012704.  Use of the Center for Nanoscale Materials, an Office of Science user facility, was supported by the U.S. Department of Energy, Office of Science, Office of Basic Energy Sciences, under Contract No. DE-AC02-06CH11357.  This work was performed in part at the University of Kentucky Center for Nanoscale Science and Engineering and Center for Advanced Materials, members of the National Nanotechnology Coordinated Infrastructure (NNCI), which is supported by the National Science Foundation (ECCS-1542164).

\noindent J.W. and X.C contributed equally to the work.

\bibliography{xray_oam.bib}

\clearpage
\onecolumngrid
\renewcommand{\thesection}{S\arabic{section}}
\renewcommand{\theequation}{S\arabic{equation}}
\renewcommand{\thefigure}{S\arabic{figure}}
\setcounter{figure}{0}
\setcounter{equation}{0}
\pagestyle{empty}

\vspace*{40pt}
\begin{center}
  \LARGE Switchable X-ray Orbital Angular Momentum \\from an Artificial Spin Ice: Supplemental Material \par
  \vspace*{3em}
  \large Justin Woods, Xiaoqian Chen, Rajesh V. Chopdekar,\\
    Barry Farmer, Claudio Mazzoli, Roland Koch, \\ Anton Tremsin, Wen Hu, Andreas Scholl,\\
    Steve Kevan, Stuart Wilkins, Wai-Kwong Kwok,\\
    Lance E. De Long, Sujoy Roy and J. Todd Hastings
\end{center}  

\section{S1. Edge dislocations in the square lattice and Sample Fabrication}

An edge dislocation can be introduced into a 2D lattice by displacing the lattice points along one Cartesian axis by a factor proportional to the azimuthal angle\cite{wilson_note_1950,willis_optical_1957}.  When an integer number of dislocations, $\mathbb{N}$, is introduced at the origin, then the lattice points satisfy
\begin{equation}
    n_1 a + \mathbb{N} a \frac{\psi}{2\pi}= \rho \cos{\psi}
    \label{eq:latticex}
\end{equation}
\begin{equation}
    n_2 a = \rho \sin{\psi}
    \label{eq:latticey}
\end{equation}
where $a$ is the lattice constant, $\psi$ and $\rho$ are the respective azimuthal and radial coordinates in the lattice plane, $n_1$  and $n_2$  are integers, and $\mathbb{N}=\norm*{\vec{t}}/a$, where $\norm*{\vec{t}}$ is the magnitude of the Burgers vector. For the structural lattice, the defect will have topological charge equal to the number of dislocations, $\mathbb{Z}_{s}=\mathbb{N}$.  If the lattice orders antiferromagnetically with twice the structural period, then the defect will exhibit half the topological charge, $\mathbb{Z}_{m}=\mathbb{N}/2$. In this work, $\mathbb{N}=2$ for all structures studied.  

The lattice is naturally distorted around the dislocation, and a strategy must be developed to accommodate this.  One approach would be to maintain the geometry of the nanomagnets themselves, and thus alter their interactions near the vertices.  A second approach would be to retain the vertex geometry by lengthening the nanomagnets, and thus increase their volume and the energy required to switch states.  Either approach is likely to alter the thermal fluctuation of nanomagnets near the defect.  We chose the later approach to keep the vertex geometry as close to the defect free system as possible.  This approach allowed the ASI to reach a signle-domain AF ground state.  

Patterns with a double dislocations were generated in Python to impart the desired topological defect.  Samples were prepared using electron beam lithography with a bilayer of PMMA positive resist on a Si substrate.  Electron beam evaporation was then used to deposit approximately 2.4 nm of Py (Ni$_{0.80}$Fe$_{0.20}$) on the substrate.  The permalloy was capped with 1.5 nm of Al to passivate the nano-islands.  Py and Al were deposited at a rate of 0.2 {\AA}/s and a pressure of $10^{-7}$ Torr.  As discussed above, the islands' dimensions vary slightly above and below the defect due to the phase progression programmed into the sample.  Patterned areas were \SI{10}{\micro\metre} $\times$  \SI{10}{\micro\metre}.

\section{S2. Vortex beams from a square artificial spin ice with a topological defect}

Here we provide additional detail about the X-ray scattering from square, magnetic lattices with topological defects.  In general the scattering amplitude is given by
\begin{align}
    G &= F(\hbar\omega,\vec{\epsilon},\vec{\epsilon'})\cdot L(\vec{Q})
\end{align}
where $F$ is the structure factor for the unit cell, $L$ is the lattice sum, and $\vec{Q}=\vec{k'}-\vec{k}$ is the scattering vector. $\vec{\epsilon}$ and $\vec{\epsilon}'$ are the polarization vectors for the incident and scattered waves respectively.  

For the lattice described by the sum in Eq. \ref{eq:latsum}, $L$ can be approximated near a reciprocal lattice vector, $\vec{Q_0}$, by\cite{wilson_note_1950,willis_optical_1957}
\begin{align}
    L'\left(\rho', \phi'\right)&=\int_{0}^{r_0}\int_{0}^{2\pi} \exp{\left(i 2\pi \rho' r \cos\left(\psi-\phi'\right)+i \vec{Q_0}\cdot \vec{t} \frac{\psi}{2\pi}\right)}r dr d\psi \label{eq:lsint}\\
    &= i^\ell \exp{(i\ell\phi'})s^{-2}\int^s_0 \xi J_\ell\left(\xi\right)d\xi \\
    &= \displaystyle i^\ell \exp(i\ell\phi') U\left(\ell,\rho'\right) \label{eq:latticeapprox}
\end{align}
where $r_0$ is the radial extent of the crystal, $s=2\pi \rho' r_0$, and $\rho'$ and $\phi'$ are polar coordinates in reciprocal space with their origins centered at the reciprocal lattice point of interest.  The integral in Eq. \ref{eq:lsint} involves  Bessel function of the first kind, $J_\ell(\xi)$, and yields a purely real, radially dependent amplitude, $U(\ell,\rho')$.  As noted in the main test, the integer $\ell$ is of critical importance and is given by
\begin{align}
    \ell=\frac{\vec{Q_0}\cdot\vec{t}}{2\pi}.  
\end{align}

The amplitude $U\left(\ell,\rho'\right)$ is given by
\begin{equation}
    U(\ell,\rho') = \frac{s^\ell{{}}_1{\mathrm{F}}_2\left(\frac{\ell}{2}+1;\ \frac{\ell}{2}+2,\ell+1;\ -\frac{s^2}{4}\right)}
    {{2^\ell\,\Gamma \left(\ell+1\right)\,\left(\ell+2\right)}}
\end{equation}
where $\mathrm{F}$ is the generalized hypergeometric function and $\Gamma$ is the gamma function.  This allows us to determine, within the validity of the approximation in Eq. \ref{eq:lsint}, that beams with $\ell\neq0$, i.e. those that carry OAM, have zero intensity at the reciprocal lattice points and are thus vortex beams.  Similarly, beams with $\ell=0$, which carry no OAM, have finite intensity at the reciprocal lattice points and are not vortex beams.  Specifically, at the reciprocal lattice vector, $\rho'\rightarrow 0$, and
\begin{equation}
    \lim_{\rho' \rightarrow 0} (L'\cdot L'^*) =
    \begin{cases}
      1/4, & \text{if}\ \ell=0 \\
      0, & \text{if}\ \ell\neq0.
    \end{cases}
\end{equation}
$U$ also indicates that the diameters of the OAM beams increase with $H$ which is consistent with increasing OAM quantum number and with the experimental results presented here. 

\section{S3. Vortex Beam Polarization}
Magnetic scattering also provides a means to control the polarization of the OAM beams.  For an incident $\sigma$-polarized beam, the structural Bragg peak appears in the $\sigma \rightarrow \sigma$ scattering channel while the pure magnetic scattering at the AF peak proceeds through the $\sigma \rightarrow \pi$ channel.\cite{hill_x-ray_1996}  As we noted in the main text, this leads to simultaneous generation of soft X-ray beams carrying even OAM with $\sigma$-polarization and beams carrying odd OAM with $\pi$-polarization. For an incident $\pi$-polarized beam, the structural Bragg peak appears in the $\pi \rightarrow \pi$ channel.  However, the magnetic scattering has both $\pi \rightarrow \sigma$ and a weaker $\pi \rightarrow \pi$ contribution.  Thus, for $\pi$-polarized incident beams the AF lattice creates elliptically polarized beams at the magnetic Bragg condition.  

The polarization sensitivity of the magnetic scattering described above provides a means to assess the phase progression of OAM beams.  Phase measurements typically require interferometry with a separate reference beam and are commonly employed for OAM beams in the visible to near-infrared spectral regions.\cite{white_interferometric_1991,pan_measuring_2018} For the ASI considered here, circularly polarized illumination naturally allows the interference of charge-scattered light in the $\sigma \rightarrow \sigma$ channel with magnetically scattered light in the $\pi \rightarrow \sigma$ channel, and vice-versa.  This phenomenon allows us to directly identify the odd OAM quantum number of the soft X-ray photons at the AF Bragg condition without a reference beam.  Of course, the contrast of the interference pattern depends on the relative intensities of the charge and magnetically scattered X-rays in the vortex regions.  

\section{S4. Verification of Results using a Sample with a Lower Transition Temperature}

\begin{figure}
    \includegraphics[width=0.9\textwidth]{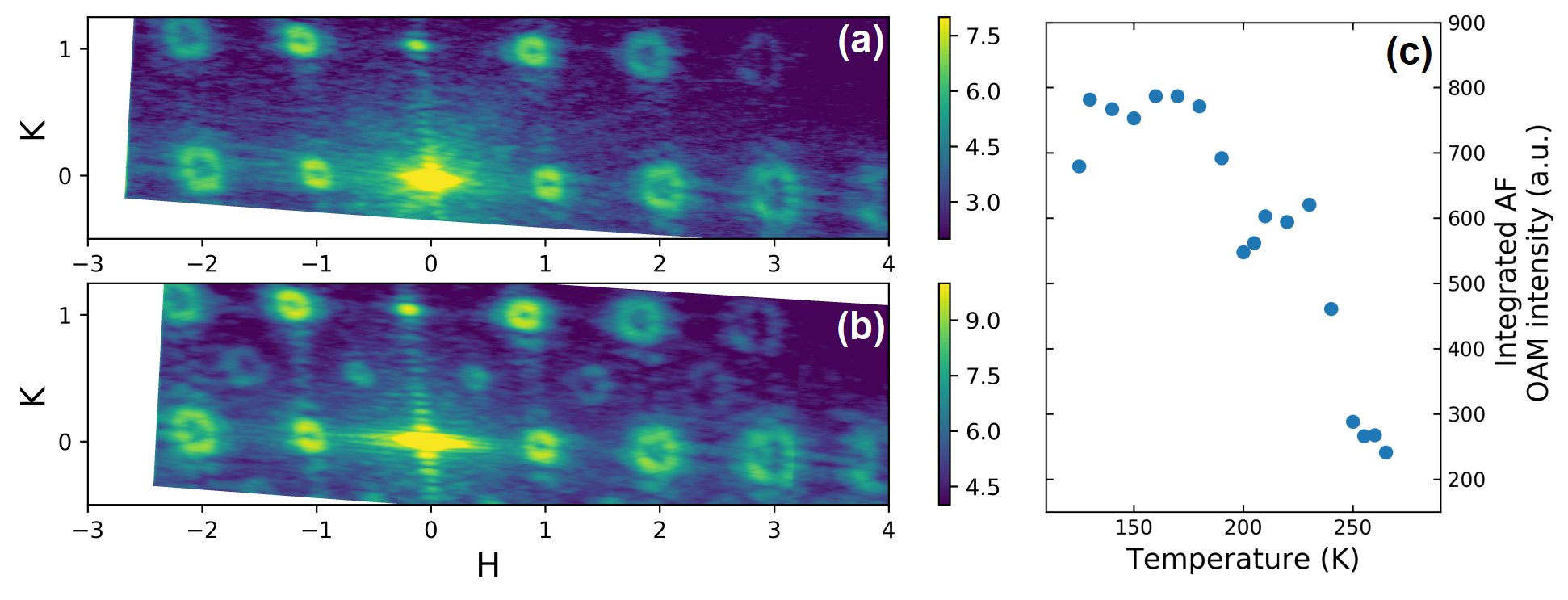}
    \caption{X-ray diffraction from a second square artificial lattice with a double edge dislocation.  The in-plane dimensions were the same as the sample described in the main text.  (a) Off-resonance (690 eV) scattered intensity shows the same vortex beams at the structural Bragg conditions.  (b) On-resonance (Fe L-edge, 708 eV) scattering reveals the additional vortex beams at the antiferromagnetic Bragg conditions, and indicates ordering in the antiferromagnetic ground state.  (c) Temperature dependence of integrated intensity of the OAM beam ($\ell=1$) near the AF Bragg condition shows the Ising-like AF-to-PM phase transition occurs near 280 K.}  
    \label{fig:COSMIC}
\end{figure}

We replicated the experiments described in the main text using a second sample with a lower antiferromagnetic-to-paramagnetic transition temperature.  The sample contained a square lattice having the same in-plane dimensions and a double dislocation.  Data was acquired at the COSMIC beamline of the Advanced Light Source using a novel Timepix-based detector \cite{tremsin_overview_2020,tremsin_optimization_2018,fiorini_single-photon_2018} and $\sigma$-polarized light.  This detector can rapidly count single X-ray photons scattered from the sample.  As shown in Fig. \ref{fig:COSMIC} (a) and (b), this sample also ordered into the antiferromagnetic ground state and produced vortex beams at both the charge and magnetic Bragg conditions.  The integrated intensity near the antiferromagnetic Bragg condition decreased as temperature increased, as shown in Fig. \ref{fig:COSMIC}(c), with a transition temperature of $T_N \approx$ 280 K.  The reduction in transition temperature, compared to the sample described in the main text, is likely the result of a slightly thinner Py film.    

\section{S5. Relaxation to the Antiferromagnetic Ground State}

After application and removal of an in-plane magnetic field, the sample relaxes to the AF ground state as show in Fig. \ref{fig:temp_field}.  This process is shown in more detail in the accompanying online video.  The video spans 510 s and is accelerated by $25.5 \times$ compared to real-time.  At the beginning of the video, within seconds of removing the magnetic field, a speckle pattern is visible near the AF Bragg condition indicating the presence of disordered AF domains.  This speckle pattern fluctuates until a vortex beam stabilizes at the mid-point of the video.  Afterwards, thermal fluctuations slightly modulate the shape of the beam.  

\end{document}